# Accuracy assessment of the UT1 prediction method based on 100-year series analysis

## Z. Malkin[1], V. Tissen[2], A. Tolstikov[2]

[1]Pulkovo Observatory, St. Petersburg, Russia
[2]Siberian Scientific Research Institute of Metrology (SNIIM), Novosibirsk, Russia

**Abstract.** A new method has been developed at the Siberian Research Institute of Metrology (SNIIM) for highly accurate prediction of UT1 and Pole coordinates. The method is based on construction of a general polyharmonic model of the variations of the Earth rotation parameters using all the data available for the last 80–100 years, and modified autoregression technique. In this presentation, a detailed comparison was made of real-time UT1 predictions computed making use of this method in 2006–2010 with simultaneous predictions computed at the International Earth Rotation and Reference Systems Service (IERS). Obtained results have shown that proposed method provides better accuracy at different prediction lengths.

**Introduction**

Requirements to prediction of the Earth rotation parameters (ERP), Universal Time (UT1) and polar motion (PM) are very high nowadays. The most exacting applications, such as navigation satellite sytems, require real-time and predicted ERP with utmost accuracy. Since international and national services provide ERP values with some delay, highly accurate 1-2-days prediction is needed for operational applications. Other applications require prediciton with greater length. For this reason, ERP prediction technique is being permanently developed.

During last several years, a new method for prediction of the Earth rotation parameters (ERP) has being developed in the Siberian Scientific Research Institute of Metrology (SNIIM) [1]. The main distinctive of this method is making use of long-time ERP series, up to 100 years, to improve the accuracy of modeling and extrapolation of the trend component of the UT1 series in the first place. The trend component is expanded in a polyharmonic time series consisting of 20 and more terms with periods from several months to several decades [2, 3]. After removing trend, the residuals are predicted making use of a modified autoregression technique. The method is described in more detail in [3–4].

In 2006, a project was started for rigorous comparison of the SNIIM prediction method with method used by International Earth Rotation and Reference Systems Service (IERS) Rapid Service/Prediction Cetre, located at the U.S. Naval Observatory (USNO). In the framework of this project, the real time predictions made simultaneously at SNIIM and USNO using the same combined USNO series updated daily were stored in real time for further processing.

Preliminary results of investigation of the accuracy of SNIIM method for UT1 prediciton was made in [2-6] for several 1–2 years time intervals. In this paper, for more reliable and detailed investigation, we have performed for the first time a comparison of all collected real-time UT1 predictions computed in 2006–2010.

**Comparison of SNIIM and USNO predicitons**

The SNIIM method for ERP prediction has been tested by means of comparison with the predictions made by the IERS Rapid Service/Prediction Center at USNO. To provide such a comparison, operational UT1 series computed daily at USNO were extrapolated at SNIIM. In total, 526 predictions made from January 2006 till December 2010 were used for our analysis. The length of compared predictions was 90 days, which corresponds to USNO daily data. The frequency of predictions gradually increased from about twice a month in 2006 to daily from September 2010 (the beginning of the IERS EOPCPPP campaign[1]). SNIIM predictions were stored together with the IERS ones made on the same day.

Thus we collected a set of 526 pairs of predictions computed simultaneously at SNIIM and USNO using the same observed EOP series. Afterwards these predictions were compared to the final USNO EOP series, and prediction errors were computed. As an accuracy estimate we computed two statistics. The first one is the root mean square error computed by

$$RMS_k = \sqrt{\frac{(x_{k,i} - x_0)^2}{N}}, \quad (1)$$

где k is the prediction length, $x_{k,i}$ is predicted ERP value in $i$-th prediction for the length $k$, $x_0$ is the final ERP value, $N$ is the number of predictions.

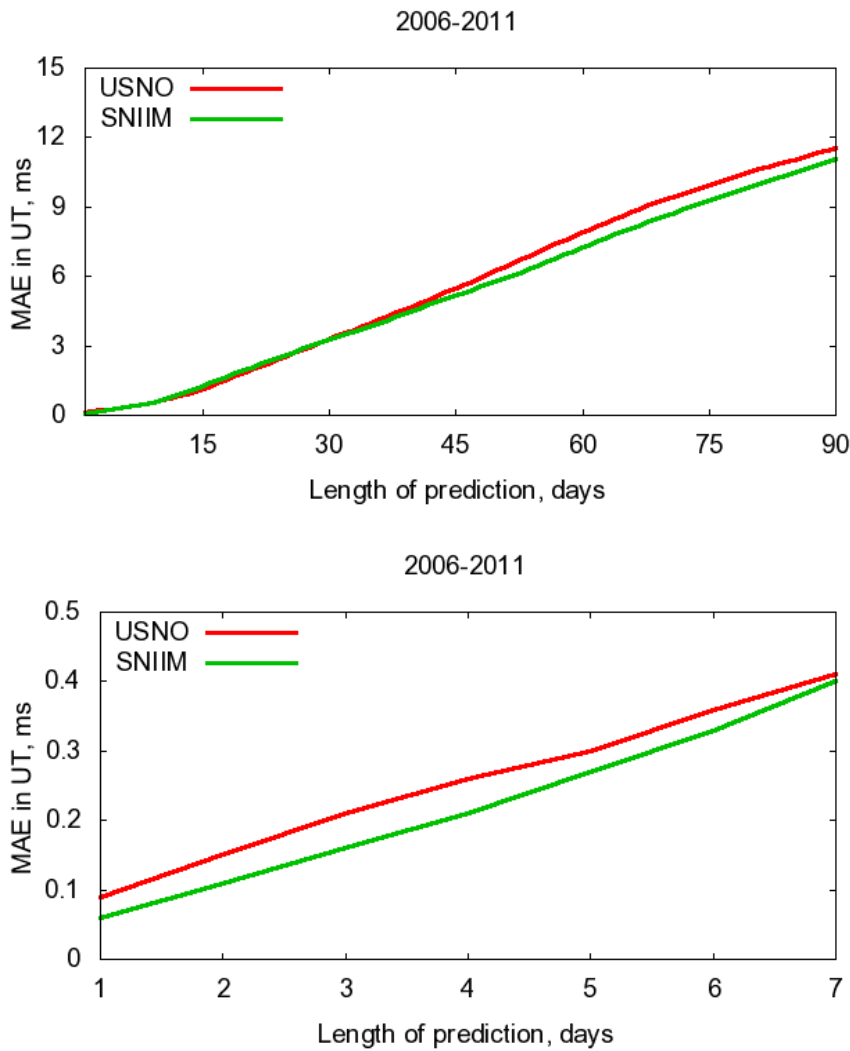

Fig. 1. Mean absolute error (MAE) of UT1 predictions made in 2006–2010 for the prediction length up to 90 (top) and up to 7 (bottom) days.

---

[1] http://maia.usno.navy.mil/eopcppp/eopcppp.html

The mean absolute error (MAE) was used in IERS prediction comparison campaign in 2006-2008 [7]. It is computed by

$$MAE_k = \frac{\sum |x_{k,i} - x_0|}{N}. \qquad (2)$$

Results are depicted in Fig. 1. RMS results are similar to MAE. One can see that SNIIM method has clear advantage for both ultra-short-term prediction of the length up to one week and medium-term predicition of the length up to 90 days. The former seems to be the most important for real-time and near-real time applications such as satellite and space positioning and navigation. Detailed results for ultra-short prediction, the most important for practice, are shown in Table 1 for whole interval and in Fig. 2 by years.

One can see from Fig. 2 that the improvement in ultra-short-term prediction remains stable with time and can be observed every year during the test period of 2006–2010. It is also interesting to see how the error of ultra-short-term prediction becomes smaller with time for both SNIIM and USNO data. It can be, most probably, explained by substantial improvement in the IERS operational UT1 series during last years [8].

Table 1. Accuracy of ultra-short-term prediction

| Prediction length, days | Prediciton error, ms | | | |
|---|---|---|---|---|
| | RMS | | MAE | |
| | USNO | SNIIM | USNO | SNIIM |
| 1 | 0.12 | 0.09 | 0.09 | 0.06 |
| 2 | 0.19 | 0.14 | 0.15 | 0.11 |
| 3 | 0.26 | 0.20 | 0.21 | 0.16 |
| 4 | 0.32 | 0.27 | 0.26 | 0.21 |

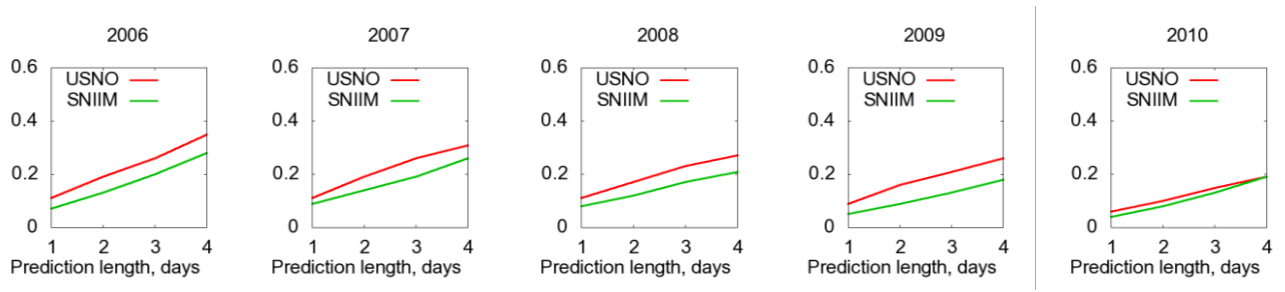

Fig. 2. Mean absolute error (MAE) of ultra-short-term UT1 predictions for testing period 2006–2010 by years, ms.

**Summary**

In this paper, the method of UT1 prediction developed at SNIIM has been tested over a 5-year interval. For this purpose, 526 90-day predictons made simultaneously at SNIIM and USNO during 2006-2010 were compared. Results of this comparison have shown that the SNIIM method has better accuracy both for ultra-short-term (up to a week) and medium-term (up to three months) predictions.

Especially important for the most exacting real-ime and near-real-time applications such as Global Navigation Satellite Systems (GNS) is substantial improvement in prediction for 1-3 days. This improvement was stable in time during 2006–2010.

**References**

1. Balahnenko A.Yu., Tissen V.M., Tolstikov A.S. Metodika prognozirovaniya DUT1 (A method for UT1 prediction). // All-Russian Conference Fundamental and Applied Time and Positioning Systems. Extended abstracts. St Petersburg: IAA RAS. 2005. P. 116–117 (in Russian).
2. Tissen V.M., Tolstikov A.S., Malkin Z.M. Neravnomernosti vrashcheniya Zemli i resultaty, dostignutye v ih prognozirovanii (Irregularities in Earth rotation and achieved results in its prediction). // Trudy Vserossiiskoi astrometricheskoi konfrentsii "Pulkovo-2009". Izv. GAO v Pulkove (Communications of Pulkovo Observatory). 2009. No 219. Issue 4. P. 329–334 (in Russian).
3. Tissen V., Tolstikov A., Malkin Z. UT1 prediction based on long-time series analysis. // Artificial Satellites, 2010, V. 45, No 2, P. 111–118.
4. Tissen V., Tolstikov A., Balahnenko A., Malkin Z. High precision prediction of universal time based on 100-year data. // Measurement Techniques, 2009, V. 52, No 2, P. 1249–1255.
5. Tissen V.M., Tolstikov A.S., Malkin Z.M. Opyt kratkosrochnogo i dolgosrochnogo prognoirovaniya parametrov vrashcheniya Zemli (Experience with short-term and long-term prediction of the Earth rotation parameters). // Geo-Sibir–2007: Proc. III Int. Congress, SGGA, Novosibirsk (2007), V. 4, Part 2, P. 92–95 (in Russian).
6. Tissen V.M., Tolstikov A.S., Malkin Z.M. Rezultaty vysokotochnogo prognozirovaniya popravki chasov dUT1 v 2008-2009 gg. dlya tselei EVO GLONASS (Results of high accuracy prediction of dUT1 in 2008-2009 for ephemerides and time support of GLONASS). // Trudy IPA RAN (IAA RAS Transactions). St. Petersburg. 2009. V. 20. P. 245–249 (in Russian).
7. Kalarus M., Kosek W, Schuh H. Current results of the Earth orientation parameters prediction comparson campaign. // Proc. Journées 2007 "The Celestial Reference Frame for the Future", Paris, France, 17–19 Sep, Ed. N. Capitaine, Paris: Observatoire de Paris, 2008. P. 159–162.
8. Improved UT1 predictions through low-latency VLBI observations. // J. of Geod. 2010. V. 84, No 6, P. 399–402.